\documentclass[conference]{IEEEtran}
\IEEEoverridecommandlockouts
\usepackage{cite}
\usepackage{amsmath,amssymb,amsfonts}
\usepackage{hyperref}
\usepackage{graphicx}
\usepackage{float} 

\begin{document}

\title{Evaluating Fault Tolerance and Scalability in Distributed File Systems: A Case Study of GFS, HDFS, and MinIO}
\author{
\IEEEauthorblockN{Shubham Malhotra}
\IEEEauthorblockA{
    Rochester Institute of Technology\\
    Dept. of Software Engineering\\
    \texttt{shubham.malhotra28@gmail.com}
}
\and
\IEEEauthorblockN{Fnu Yashu}
\IEEEauthorblockA{
    Stony Brook University\\
    Dept. of Computer Science\\
    \texttt{yyashu@cs.stonybrook.edu}
}
\and
\IEEEauthorblockN{Muhammad Saqib}
\IEEEauthorblockA{
    Texas Tech University\\
    Dept. of Computer Science\\
    \texttt{saqibraopk@hotmail.com}
}
\and
\IEEEauthorblockN{Dipkumar Mehta}
\IEEEauthorblockA{
    C.K.Pithawalla College of Eng. and Tech.\\
    \texttt{dipkumar.mehta@gmail.com}
}
\and
\IEEEauthorblockN{Jagdish Jangid}
\IEEEauthorblockA{
    Infinera Corp\\
    \texttt{jangid.jagdish@gmail.com}
}
\and
\IEEEauthorblockN{Sachin Dixit}
\IEEEauthorblockA{
    Stripe Inc.\\
    \texttt{spdixit@gmail.com}
}
}

\maketitle

\begin{abstract}

Distributed File Systems (DFS) are essential for managing vast datasets across multiple servers, offering benefits in scalability, fault tolerance, and data accessibility. This paper presents a comprehensive evaluation of three prominent DFSs—Google File System (GFS), Hadoop Distributed File System (HDFS), and MinIO—focusing on their fault tolerance mechanisms and scalability under varying data loads and client demands. Through detailed analysis, how these systems handle data redundancy, server failures, and client access protocols, ensuring reliability in dynamic, large-scale environments is assessed. In addition, the impact of system design on performance, particularly in distributed cloud and computing architectures is assessed. By comparing the strengths and limitations of each DFS, the paper provides practical insights for selecting the most appropriate system for different enterprise needs, from high-availability storage to big data analytics.

\end{abstract}

\section{Introduction}
\label{sec: intro}

A distributed file system (DFS) is a file system implemented on a client/server architecture, where one or more central servers store files that can be accessed by multiple remote clients, which is based on the access protocol defined by the server and the level of access granted to the client. The fundamental feature of a distributed file system is that the interface of the file access by the remote client is as if the file is being accessed from their local machine. Therefore, files can be cached, accessed, and managed on the local client machine, but the process is managed by a single, or a network of centralized servers.

DFSs constitute the primary support for data management. They provide an interface whereby to store information in the form of files and later access them for read and write operations. Among the several implementations of file systems, few of them specifically address the management of huge quantities of data on a large number of nodes. Mostly, these file systems constitute the data storage support for large computing clusters, supercomputers, massively parallel architectures, and lately, storage/computing clouds \cite{buyya2013mastering}.

Distributed file systems allow for efficient, easily manageable, and extensible data storage and sharing given a network and relevant network protocols. As mentioned before, one of the key features of a DFS is the ability of the server to set up a protocol to restrict client access based. This allows multiple clients to be able to access multiple files hosted on a server with varying degrees of access. DFSs depend on three main notions of \textit{transparency, fault tolerance}, and \textit{scalability}.

\subsection{Transparency}
\cite{floyd1989transparency} implies the user should be able to access the distributed file system regardless of their login node or client machine and, based on their access level, should be able to perform the same operations without caring about the faults in the system because of the fault-tolerant mechanisms of the distributed system itself. Therefore, the client should be able to access the requisite files without worrying about consistency, faults, and the complexity of the underlying file system.

\subsection{Fault Tolerance}
\cite{becker1994application} should not be stopped in case of transient or partial failures. Faults could be of network failure or server failure, which would result in data and service unavailability, compromising data integrity. Fault tolerance plays an integral role in consistency when several users concurrently access the data, typically when the data is stored in multiple server locations. The cost and complexity of designing such systems increases significantly based on the severity of the failures and the relative importance of the data to the server host(s).

\subsection{Scalability}
\cite{mccreadie2012mapreduce} is the ability to efficiently leverage large amounts of servers, which can be either dynamically or continuously added to the system. A distributed file system should be scalable to account for maintaining replicas and increasing fault tolerance as the number of files, size of files, or number of clients increase. Scalability implies both storage space as well as distributed compute. Some systems that adopt a centralized architecture provide tools such as multi-threading for scaling client access to files in DFSs.

\subsection{Terms and Jargon used in Distributed File System Literature}

The literature on DFSs is based on a combination of literature in distributed systems as well as local file system and computer system architecture. Therefore, a high density of jargon has evolved that is specific to this field. A few of these terms have been detailed here.

\subsubsection{High Availability}
High availability is that characteristic of a system, which aims to ensure an agreed level of operational performance, usually uptime, for a higher than normal period. This implies that a highly available system will:
\begin{itemize}
    \item \textbf{Add redundancy} to remove single points of failure. One component failing shouldn't bring the system down.
    \item \textbf{Reliably crossover}. Given that the system is redundant, the crossover point is prevented from becoming a single point of failure.
    \item \textbf{Limit the chances of failure}. Given the above two principles, failures shouldn't occur. If and when they do, they should be detected as soon as possible, and regular maintenance must be done.
\end{itemize}

\subsubsection{Shards}
A database shard is a horizontal partition of data in a database or search engine. Each individual partition is referred to as a shard or database shard. Each shard is held on a separate database server instance to spread load.

The advantages of having shards include:
\begin{itemize}
    \item Total rows per table reduced
    \item Index size reduced, improving search performance
    \item Distribute database over more hardware.
    \item Some data is naturally distributed like this (e.g., country-specific data stored in a shard of database that is physically present in that country for lower network latency)
\end{itemize}

However, there are some disadvantages as well, which are:
\begin{itemize}
    \item Increased complexity of SQL
    \item Sharding introduces complexity
    \item Single point of failure
    \item Failover servers more complex
    \item Backups more complex
    \item Operational complexity added
\end{itemize}

\subsubsection{Volumes}
\subsubsection{Volumes}
Volumes serve as a management construct that logically organizes a cluster’s data. Since a container is always associated with one volume, all replicas of that container are also tied to the same volume. Volumes do not have a predefined size and consume disk space only when the MapR file system stores data within a container assigned to the volume. A large volume may consist of anywhere from 50 to 100 million containers. Common use cases include creating volumes for specific users, projects, or different stages, such as development and production environments. For instance, when an administrator needs to arrange data for a particular initiative, a unique volume can be created for that purpose. The MapR file system will then group all containers that hold the data related to the initiative within the designated volume. A cluster can host multiple volumes.

Mirror volumes are non-writable replicas of a primary volume. These mirror volumes can be promoted to writable volumes. This functionality is particularly useful in disaster-recovery situations, where a read-only mirror can be promoted to a writable volume to serve as the primary data storage. Furthermore, writable volumes that have been mirrored can also be converted into mirrors (to establish a reverse mirroring relationship). Writable volumes can additionally be reverted to read-only mirrors.

\subsubsection{Snapshots}
Snapshots allow you to revert to a previously saved and stable data set. A snapshot is a non-writable replica of a volume at a specific point in time, offering recovery to that exact moment. Snapshots only track the changes made to the volume's data, which results in an efficient use of disk resources within the cluster. They help preserve access to past data and safeguard the cluster against user or application mistakes. Snapshots can be created manually, or you can automate the process with scheduled tasks.

\subsubsection{MapReduce}
MapReduce is a computational framework that enables applications to process extensive amounts of data. It operates by executing applications concurrently across a group of inexpensive machines in a dependable and fault-tolerant manner. The framework consists of several map and reduce tasks, with each task processing a portion of the data. This enables the workload to be distributed throughout the cluster. The map tasks are responsible for loading, parsing, transforming, and filtering the data. On the other hand, reduce tasks aggregate and group the intermediate results produced by the map tasks.

The input file for a MapReduce job resides on HDFS. The input format determines how the file is divided into smaller chunks, known as input splits. These input splits provide a byte-level representation of a segment of the input file. The map task then processes the input split locally on the node where the data is stored. This eliminates the need for data transfer over the network, ensuring that the data is processed locally.

\subsection{Types of Distributed File Systems}

By architecture design, DFSs can be classified as \textbf{Client-Server Architectures} and \textbf{Cluster-Based Distributed File System}. 
\begin{itemize}
    \item A client-server architecture has several servers that manage the data store and caching functions, access management, and data transfer between different clients. The data transfer and the metadata are not decoupled, meaning that the data access is based on a global namespace, which is shared by multiple clients. In a client-server architecture, the addition of new servers increases processing capacity as well as storage.
    
    \item On the contrary, a cluster-based distributed file system decouples the metadata and the data transfer, by having dedicated servers to manage storage and others to manage metadata. Increasing the number of storage servers increases the overall capacity of the distributed system, but might negatively affect the query time if the metadata server(s) are not increased in capacity. Systems having a single metadata server are referred to as centralized cluster-based DFSs and often have a single point of failure. Distributed metadata servers occur in totally distributed cluster-based DFSs.
\end{itemize}

In terms of cache consistency, the distributed systems can be analyzed as systems that implement \textit{Write Once Read Many (WORM)}, \textit{Transactional Locking}, and \textit{Leasing} \cite{reed1996analysis}. 

\begin{itemize}
    \item \textbf{Write Once Read Many (WORM)} is an approach that ensures mutual consistency across the servers and clients based on a single write to a file segment which can then be accessed multiple times for reading. Once a file is created, it cannot be modified, rewritten, or appended. Cached files are in read-only mode, therefore each read reflects the latest version of the data. Consistency here mirrors guaranteed eventual consistency in distributed systems as the final version of a

\end{itemize}

\section{Hadoop Distributed File System (HDFS)}

The Hadoop Distributed File System (HDFS) \cite{borthakur2007hadoop} is designed to run on commodity hardware, making it extensible and easily accessible for consumer-grade networks and low-performance servers. While similar to other distributed file systems in its WORM model and persistent failure handling, HDFS stands out for its high fault tolerance and cost-effectiveness, as it is optimized for deployment on inexpensive hardware.

HDFS follows a centralized architecture where metadata is managed by a single server. This server maintains a persistent copy of metadata, enabling quick restart and migration of metadata from the secondary to the primary data node when necessary.

HDFS provides high throughput for large data sets and is ideal for applications requiring streaming access. It was originally developed for the Apache Nutch web search engine project and is part of the Apache Hadoop Core initiative. HDFS divides files into large blocks, distributing them across cluster nodes, and runs packaged code on these nodes to process data in parallel. This method benefits from data locality, improving processing speed and efficiency compared to traditional supercomputing models that rely on high-speed networking to distribute computation and data.

The Hadoop architecture is open-source and available at \url{http://hadoop.apache.org/}.

\subsection{Goals of HDFS}

HDFS is designed for deployment across hundreds of server nodes with attached data stores. Its core goals include:

\begin{itemize}
    \item \textbf{Efficient Fault Detection and Recovery}: Given that hardware failure is common in large systems, HDFS is built to detect faults quickly and recover automatically. With potentially thousands of nodes, fault tolerance is a primary design consideration.
    
    \item \textbf{Optimized Data Access}: HDFS is tailored for streaming data access, supporting batch processing rather than interactive use. High throughput is prioritized over low latency, with certain POSIX requirements relaxed to boost performance.
    
    \item \textbf{Handling Large Datasets}: HDFS is optimized for large-scale applications, capable of managing petabytes of data across hundreds of nodes. It supports millions of files within a single instance while maintaining high aggregate bandwidth.
    
    \item \textbf{Simplified Coherency Model}: HDFS adopts a write-once-read-many (WORM) model, where files are not modified once closed. This simplifies data consistency issues, enabling efficient data access and making HDFS ideal for batch-processing applications like MapReduce.
    
    \item \textbf{Cross-Platform Portability}: HDFS is designed to be easily portable across different platforms, ensuring broad compatibility and adoption for various applications.
\end{itemize}

\subsection{\textbf{Architecture Description}}

\begin{figure}
    \centering
    \includegraphics[width=\linewidth]{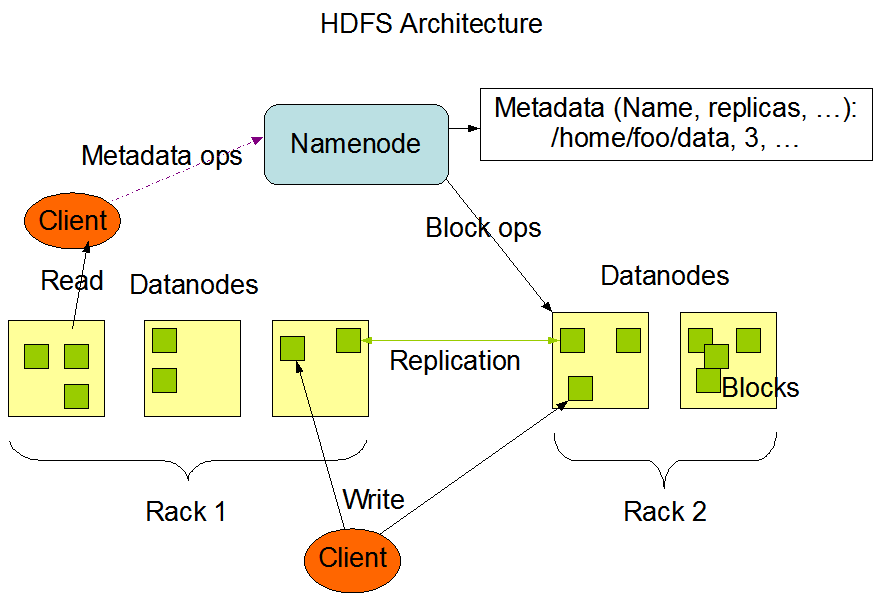}
    \caption{The Hadoop Distributed File System Architecture}
    \label{fig: hdfs}
\end{figure}

As seen in Figure \ref{fig: hdfs}, an HDFS cluster consists of a single \textbf{NameNode}, a master server that manages the file system namespace and regulates access to files by clients. Therefore, the NameNode is the master metadata and processing node. 

\subsection{HDFS Components and Architecture}

HDFS consists of two primary components: the \textbf{NameNode} and the \textbf{DataNodes}. The NameNode manages the file system namespace, handling operations like file creation, deletion, and renaming while also mapping file blocks to DataNodes. The DataNodes manage the storage of these blocks and execute tasks like block creation, deletion, and replication as directed by the NameNode. These components run on commodity machines, usually using a GNU/Linux OS. The system is implemented in Java, ensuring cross-platform compatibility.

\subsubsection{Data Replication and Reliability}

HDFS ensures high fault tolerance by replicating file blocks across multiple data nodes. The replication factor is configurable, allowing fine-tuned fault tolerance depending on the file's importance. For efficient data replication, the NameNode tracks the state of the DataNodes and manages block replication decisions. HDFS supports a WORM model where files, once written, cannot be modified except for appending or truncating.

\paragraph{Replica Placement} 
HDFS optimizes replica placement to balance reliability, availability, and network bandwidth. Data replicas are placed with respect to rack-awareness, ensuring that replicas are distributed across different racks to minimize data loss in case of a rack failure. The default policy places replicas in such a way that minimizes inter-rack traffic, improving write performance without compromising data reliability.

\paragraph{Replica Selection} 
To optimize read performance, HDFS prioritizes retrieving data from the closest replica. When the cluster spans multiple data centers, local replicas are preferred to minimize latency.

\subsubsection{Metadata Persistence and Communication Protocols}

The NameNode stores metadata using a transaction log, called the \textit{EditLog}, and an image of the file system namespace, the \textit{FsImage}. The EditLog records changes to the file system, while the FsImage represents a snapshot of the file system's state. Periodic checkpointing ensures the file system’s consistency by merging the EditLog with the FsImage. The DataNode stores blocks of data on its local file system and sends periodic reports, known as Block reports, to the NameNode, listing the blocks it manages.

HDFS communication uses TCP/IP protocols. Clients interact with the NameNode via the \textit{ClientProtocol}, and DataNodes communicate with the NameNode using the \textit{DataNode Protocol}. Both protocols are built upon Remote Procedure Calls (RPCs), where the NameNode only responds to incoming RPC requests from clients or DataNodes, never initiating them.

\subsection{\textbf{Client Operations}}

Data read requests are processed by HDFS, with interactions between the NameNode and DataNodes. The reader, referred to as a \texttt{client}, follows these steps:

\subsubsection{Read Operations in HDFS}
\begin{enumerate}
    \item The client initiates a read request by calling \verb|open()| on the \texttt{FileSystem} object, which is of type \texttt{DistributedFileSystem}.
    \item The \texttt{DistributedFileSystem} connects to the NameNode using RPC and retrieves metadata such as block locations.
    \item The NameNode returns addresses of DataNodes storing the first few blocks.
    \item The client receives an object of type \texttt{FSDataInputStream}, containing \texttt{DFSInputStream} that facilitates interactions with DataNodes and NameNodes. 
    \item The client repeatedly calls \verb|read()|, which fetches data in streams until the end of a block.
    \item Upon reaching the block's end, \texttt{DFSInputStream} closes the connection and locates the next DataNode for the subsequent block.
    \item After reading, the \verb|close()| method is invoked to finish the operation.
\end{enumerate}

\subsubsection{Write Operations in HDFS}
\begin{enumerate}
    \item The client initiates the write operation by calling \verb|create()| on the \texttt{DistributedFileSystem} to create a new file.
    \item The NameNode verifies file existence and client permissions. If valid, a new file record is created; otherwise, an \texttt{IOException} is thrown.
    \item Upon success, the client receives a \texttt{FSDataOutputStream} object for writing data.
    \item \texttt{FSDataOutputStream} contains \texttt{DFSOutputStream}, which manages communication with DataNodes and the NameNode. Data is written in packets that are queued in \texttt{DataQueue}.
    \item The \texttt{DataStreamer} consumes the packets from \texttt{DataQueue} and requests block allocation from the NameNode, choosing DataNodes for replication.
    \item A pipeline of DataNodes is established for replication. With a replication factor of 3, data is streamed from the first DataNode to the second and then to the third.
    \item \texttt{DFSOutputStream} maintains an \texttt{Ack Queue} to store packets awaiting acknowledgment from DataNodes.
    \item Once acknowledgments are received, the packet is removed from the \texttt{Ack Queue}. If any DataNode fails, the operation is retried using the remaining packets.
    \item Upon completion, the client calls \verb|close()|, flushing remaining data and waiting for final acknowledgment.
    \item The NameNode is notified once the write operation is finished.
\end{enumerate}

\subsection{\textbf{Properties of HDFS}}

\subsubsection{Robustness}
HDFS ensures data reliability despite failures, including NameNode, DataNode, and network partition failures.

\paragraph{Data Disk Failure} Periodically, each DataNode sends a heartbeat to the NameNode. In the case of a network partition, a lack of heartbeat indicates failure, causing the NameNode to mark the DataNode as dead. Blocks registered to the dead node become inaccessible, and replication is triggered to restore the desired replication factor.

\paragraph{Cluster Rebalancing} HDFS supports dynamic data movement to balance storage. Although rebalancing schemes are not yet implemented, data can be reallocated to maintain optimal space usage.

\paragraph{Metadata Disk Failure} The \texttt{FsImage} and \texttt{EditLog} are critical for HDFS operation. The NameNode can maintain multiple copies of these files to prevent data loss. Synchronous updates of these copies ensure consistency but may affect metadata transaction throughput.

\paragraph{Snapshots} Snapshots capture a point-in-time copy of data, enabling rollback to a prior state in case of corruption.

\subsubsection{Accessibility}
HDFS supports diverse access methods, including the native Java FileSystem API, a C wrapper, and REST API. A web interface and NFS gateway allow browsing and mounting of HDFS filesystems.

\subsubsection{Data Organization}
HDFS is optimized for large files and is suitable for applications requiring high-throughput, read-heavy operations. It supports the write-once-read-many model, with files typically split into 128 MB blocks distributed across DataNodes.

\paragraph{Replication Pipeline} With a replication factor of 3, data is written to a pipeline of DataNodes. Each DataNode in the pipeline stores and forwards data to the next. This pipelined approach ensures parallel processing and data redundancy.

\subsubsection{Space Reclamation}
When enabled, HDFS moves deleted files to a trash directory rather than immediately removing them. Files remain in the trash for a configurable period, and old checkpoints are deleted after expiry. After the retention period, the NameNode permanently deletes the file and frees associated blocks, though this may result in some delay in reclaiming space.

\section{Google File System (GFS)}

The Google File System (GFS) is a scalable, distributed file system designed to handle large, data-intensive applications. It ensures fault tolerance while operating on low-cost commodity hardware and delivers high aggregate performance to numerous clients. Large-scale GFS deployments can offer hundreds of terabytes of storage spread across thousands of disks across over a thousand machines, with concurrent access from hundreds of clients.

While GFS provides a familiar file system interface, it does not strictly adhere to POSIX semantics. Files are organized hierarchically in directories and identified by pathnames. Key operations such as \texttt{create}, \texttt{delete}, \texttt{open}, \texttt{close}, \texttt{read}, \texttt{write}, and especially \texttt{append} to files are supported.

\subsection{\textbf{Objectives of GFS}}

The primary goal of GFS in large-scale environments is to offer a highly transparent, consistent, and fault-resistant framework for distributed file reading, writing, accessing, and appending. This objective is achieved through a combination of principles, as detailed below.

Firstly, component failures are considered typical rather than exceptional. The system comprises hundreds or even thousands of storage machines made from affordable commodity hardware, with a similar number of client machines accessing them. Given the sheer volume and variability of these components, it is practically inevitable that some will be non-functional at any given time, and certain failures will be irreparable. The system has encountered issues stemming from application bugs, operating system failures, human mistakes, as well as failures in disks, memory, connectors, networking, and power supplies. As a result, the system must integrate continuous monitoring, error detection, fault tolerance, and automatic recovery mechanisms.

Secondly, the size of files far exceeds traditional standards. Multi-gigabyte files are commonplace. Each file often consists of numerous application objects, such as web documents. Managing billions of roughly kilobyte-sized files within fast-growing datasets of several terabytes containing billions of objects becomes unmanageable. Even if the file system could theoretically support it, this scale of data requires rethinking certain design assumptions and parameters, such as I/O operations and block sizes.

Thirdly, the majority of files are modified by appending data rather than overwriting existing content. Random writes within a file are virtually nonexistent. After a file is written, it is typically only read, and often sequentially. This access pattern is common for a variety of data types. For example, large repositories of data are often scanned by analysis programs, while other types may represent continuously generated data streams or archival information. Additionally, intermediate results may be produced on one machine and later processed on another. Due to these usage patterns with large files, the focus is placed on optimizing the performance of appending operations and ensuring atomicity, while caching data blocks on the client side become less relevant.

Fourthly, the co-design of applications and the file system API enhances overall system flexibility. For example, the consistency model of GFS has been relaxed significantly, simplifying the file system while reducing the burden on applications. Furthermore, an atomic append operation has been introduced, allowing multiple clients to append data to a file concurrently without the need for additional synchronization.
\subsection{\textbf{Architecture Description}}
\begin{figure}
    \includegraphics[width=0.5\textwidth]{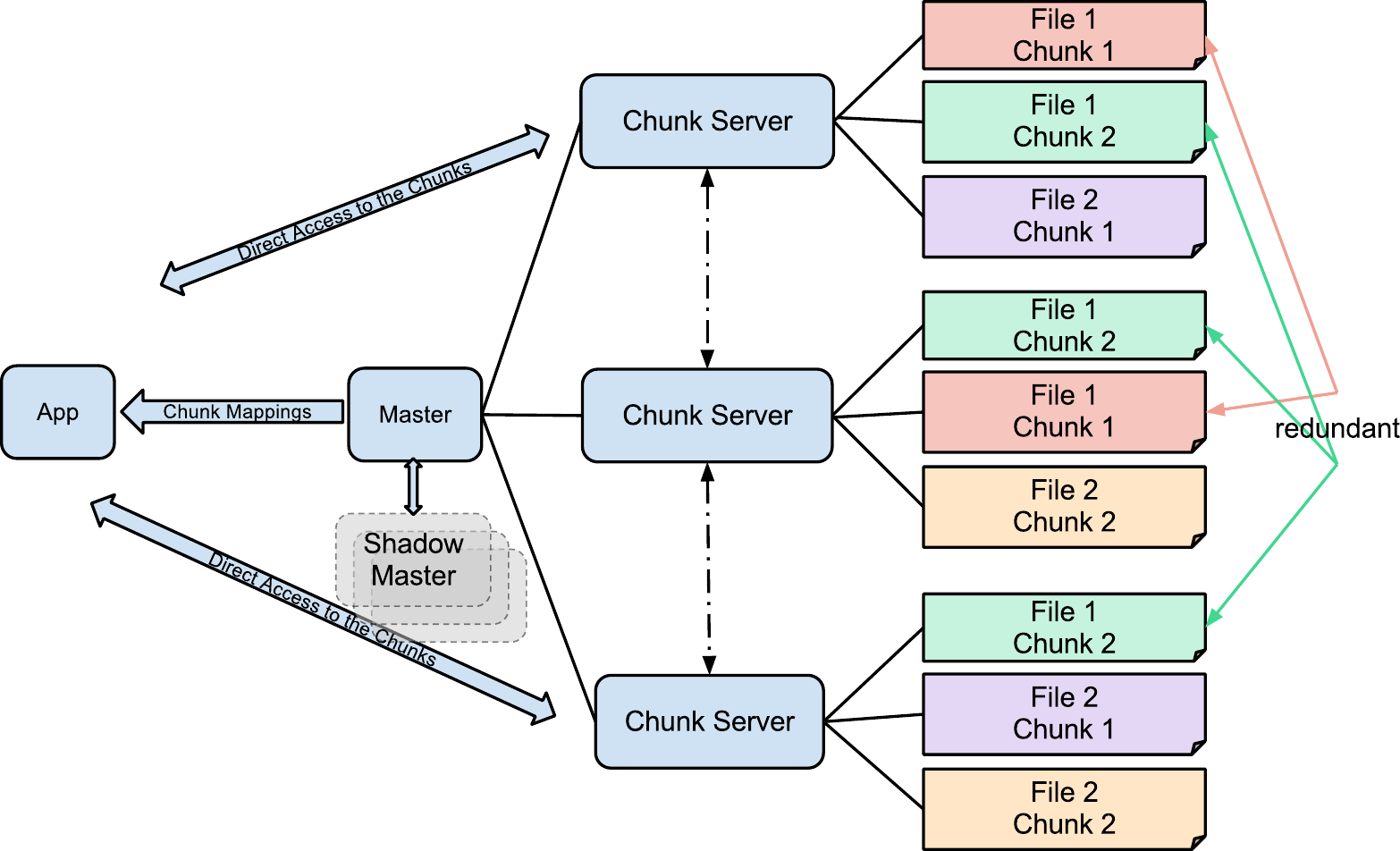}
    \caption{GFS architecture}
    \label{fig:architecture}
\end{figure}

Severe failures or network partitions. Minio ensures that the data remains safe and accessible by employing a highly resilient architecture with redundancy across multiple locations. By using distributed erasure coding and continuous replication, Minio provides excellent data durability and availability across different failure scenarios.

\begin{itemize} \item \textit{Erasure Coding}: A method for providing redundancy that minimizes the overhead compared to traditional replication. This allows Minio to store data more efficiently and with fewer resources while maintaining durability. \item \textit{Bitrot Protection}: Minio includes mechanisms to detect and repair Bitrot, ensuring that data corruption due to disk errors or other hardware failures does not go unnoticed. \item \textit{Encryption and WORM (Write Once Read Many)}: Sensitive data can be encrypted at rest and during transit to ensure security, and the WORM feature can enforce data immutability, preventing unauthorized changes. \item \textit{Identity Management}: Integrates with identity providers for secure access control and authentication, providing robust mechanisms for managing who can access the data and how. \item \textit{Global Federation}: Allows you to federate multiple Minio instances across different regions or data centers, creating a unified global object storage system for the applications. \item \textit{Multi-Cloud Support}: With Minio's gateway mode, users can seamlessly integrate Minio with existing public cloud providers, offering hybrid cloud storage solutions. \end{itemize}

Minio’s scalability makes it a competitive alternative for large-scale storage solutions, and it integrates easily with applications, including those built using cloud-native architectures. Its simplicity in deployment and operation makes it appealing for both small and large organizations seeking to manage their storage infrastructure independently.

In summary, Minio provides a robust, scalable, and cost-effective solution for distributed object storage, with the flexibility to handle demanding workloads and large data sets while supporting advanced features like encryption, federation, and erasure coding for data protection and availability.

\subsection{\textbf{Architecture Description}}
\begin{figure}
    \centering
    \includegraphics[width=\linewidth]{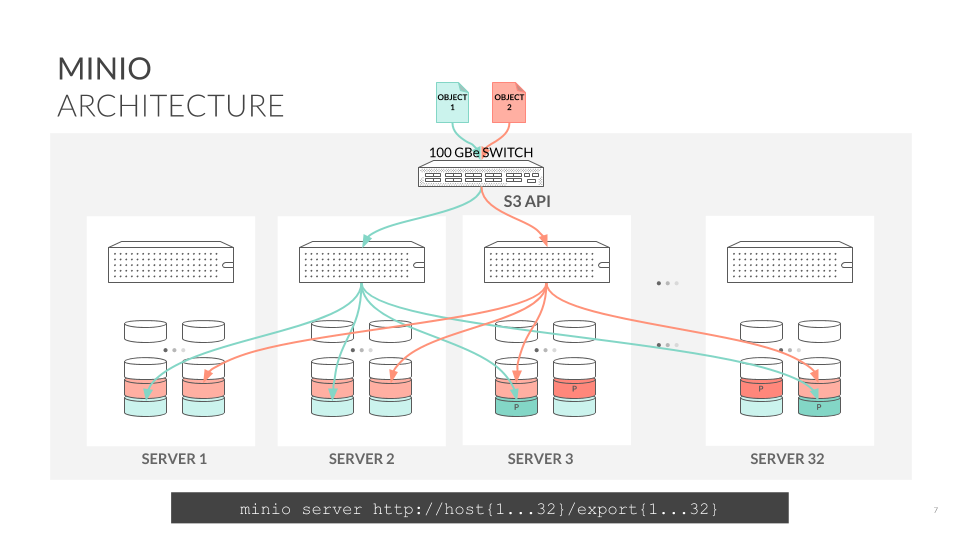}
    \caption{MinIO Architecture Diagram}
    \label{fig: minio}
\end{figure}

By design, MinIO is cloud native and can be run using lightweight containers managed by external orchestration services (ex. Kubernetes). A competitive advantage of Minio is that it fits the entire server into a single ~40MB static binary. Though not compiled at source, it is highly efficient in its use of CPU and memory resources and thus allows the co-hosting of a large number of tenants on shared hardware.

MinIO fucntions well on commodity servers (though often benchmarked with state-of-the-art hardware) with locally attached drives (JBOD/JBOF). Keeping to the norms of efficient scalability, all the servers in a cluster are equal in capability (fully symmetrical architecture). In fact, there are no name nodes or metadata servers.

An important implementation detail within MinIO is that it writes data and metadata together as objects, thus eliminating the requirement for a separate metadata database. MinIO's resiliency can also be attributed to the fact that all functions it uses are inline and strictly consistent. 

While a MinIO cluster is a collection of distributed MinIO servers with one process per node, it itself runs in the user space as a single process and uses lightweight co-routines for high concurrency. A deterministic hashing algorithm is utilized to place objects within Erasure sets, which have 16 drives per set by default.

Architecturally, MinIO is designed to operate at scale across multi-datacenter cloud services. In it, each tenant runs their own fully-isolated MinIO cluster, protecting them from all possible disruptions to to things like upgrades, updates, and security breaches. Additionally, each tenant can scale independently by federating clusters across geographical locations. Minio's Object Storage Architecture is also quite interesting. Most existing storage solutions follow a system that involves a multi-layer storage architecture comprising a durable block layer at the bottom , a filesystem as "middleware", and APIs on top implementing protocols for various operations on files, blocks, and objects.

While public cloud architectures provide separate object, file and block storage, Minio follows a fundamentally different architecture: using a single layer to achieve everything. As a result, the minio object server is high-performance and lightweight. 

\subsubsection{MinIO Design Decisions}

\paragraph{Lambda* Function Support}
Enterprise standard messaging platforms can be used to deliver notifications of events via the Amazon-compatible lambda event notifications. This allows notifications for object-level events/actions like access, creation, and deletion to be delivered to the application layer.

\paragraph{Linear Scaling}
By taking inspiration from hyperscalers, Minio clusters can be deployed at a wide range of nodes ranging from 4 to 32. Through its feature of Federation, multiple clusters are can easily be joined together under the same "global" namespace, which essentially is a single entity to the outside world. As a result of federation:
    \begin{itemize}
        \item All nodes are considered equal
        \item Any node can serve requests concurrently
        \item A DLM (Distributed Locking Manager) helps a cluster manage updates and deletions
        \item There is no performance degradation of an individual cluster due to the addition of more clusters under the global namespace
        \item Cluster-level failure domain
    \end{itemize}

\paragraph{Erasure Code}
The integrity of object data is maintained by erasure coding and bitrot protection checksums. For some background information, erasure code is a mathematical algorithm that can be used to reconstruct missing or corrupted data. To achieve this the Reed-Solomon code is used to shard objects into data and parity blocks, and hashing algorithms are used to help protect individual shards. As a result, Minio is resilient to silent data corruption (which happens quite often at a petascale of data) and other hardware failures. Raid configurations and data replicas suffer from high storage overheads. Erasure code solves this problem while allowing for the loss of up to 50 percent of drives and 50 percent of servers. RAID-6, as a case study, is only able to withstand two-drive failures. By applying erasure code to individual objects, Minio allows the healing of one object at a time. On the other hand, RAID-protected storage solutions have healing done at a RAID Volume level, which impacts performance for files within that level for the duration of the healing. 

\subsection{\textbf{Properties of MinIO Distributed File System}}
\begin{itemize}
    \item \textit{Performance}: One platform allows support for various different use cases. Industrial use cases have shown that a user can run multiple Spark*, Presto*, and Hive* queries or even deploy large AI workloads and algorithms without encountering any issues related to storage (read or write). Minio object storage allows for high throughput and low latency for cloud-native applications. Coupled with the latest hardware and network infrastructure, Minio outperforms all traditional object storage. 
    \item \textit{Scalability}: Single clusters can be federated with other clusters to create namespaces that can span multiple data centers. Incremental expansion of physical servers is what leads to the gradual expansion of the global namespace. Though a simple design, it leverages all cutting-edge knowledge of hyperscalers. 
    \item \textit{Ease of Use}: Start-up configuration time is only a few minutes, given the existence of a single 40MB binary for the server. Default configurations work very well, thus allowing for use in smaller projects as well, without full knowledge of system administration. 
    \item \textit{Encryption and WORM}: Unique keys are used to encrypt each object (per-object key). By supporting integration with external key management solutions, state-of-the-art cryptography solutions can be used to secure and manage encryption keys without having any coupling with Minio. WORM (write once, read many) mode prevents data tampering.
    \item \textit{Identity and Access Management}: Support for OpenID*-compatible identity management servers allows for secure and well-tested access control. Temporary rotating credentials within Minio prevent the need to embed long-term credentials within an application. 
    \item \textit{High Availability}: Even if a minio cluster loses up to 50\% of its drives and servers, Minio will continue to serve objects. Additionally, total rack failures are also mitigated (if the cluster is deployed across racks). These features are achieved by using Minio's distributed erasure code, which uses multiple redundant parity blocks to protect data. In case of data center level outages, there is also support for continuous mirroring to remote sites (disaster recovery). 
    \item \textit{Metadata Architecture}: As a result of not having a separate metadata store, all failures are contained within an object and do not spill over to the rest of the system. Additionally, all operations within Minio are performed atomically at an object level of granularity. Other data integrity requirements are satisfied by the existence of erasure code and bitrot hash per object. 
    \item \textit{Strict Consistency}: Strictly consistent operations allow for the system to survive crashes even in the middle of other workloads without any loss to pre-existing data. This is highly useful in use cases related to machine learning and other big data workloads.
    \item \textit{Geographic Namespace}: Using Mini's federation feature, users can opt to scale in an incremental fashion rather than having to deploy across data centers via hyperscalers from the start itself. It can be deployed in units that have a failure domain restricted to the size at which it is currently scaled.
    \item \textit{Cloud-native design}: Kubernetes and other orchestration platforms can easily be used along with Minio's multi-instance and multi-tenant design. Containerized deployment of Minio leads to the use of these orchestration services for reliable scaling. Each instance of Minio can be provisioned on demand through self-service registration. While traditional monolithic storage systems (which have its own disadvantages) do compete with Kubernetes resource management, they lack its easily scalable nature and the ability to pack many tenants simultaneously on the same shared infrastructure.
    
\end{itemize}

\section{Comparing the DFSs}

\begin{table}[htbp]
\centering
\caption{Comparison of DFS (Part 1)}
\label{tab:hdfsvsminio1}
\begin{tabular}{|c|c|c|c|}
\hline
\bf DFS & \bf Written In & \bf License & \bf Access API \\
\hline
\it Minio & Go & Apache v2 & AWS S3 API  \\
\hline
\it HDFS & Java & Apache v2 & Java and C client, HTTP  \\
\hline
\it GFS & \texttt{UNK} & Proprietary & Native File System API \\ 
\hline
\end{tabular}
\end{table}

\begin{table}[htbp]
\centering
\caption{Comparison of DFS (Part 2)}
\label{tab:hdfsvsminio2}
\begin{tabular}{|c|c|c|c|}
\hline
\bf DFS & \bf High Availability & \bf Shards & \bf Release \\
\hline
\it Minio & Yes & Yes & 2014 \\
\hline
\it HDFS & Transparent master failover & No & 2005 \\
\hline
\it GFS & Yes & Yes (by Spanner) & 2003 \\ 
\hline
\end{tabular}
\end{table}

\subsection{\textbf{HDFS vs MinIO}}

While HDFS has been a long-standing player in the distributed filesystem market, Minio has shown to outperform it in many of its seminal tasks. Their core difference lies in the philosophy of storage. HDFS achieves its high throughput values by colocating compute and data on the same nodes. As a result they get to exploit fewer network calls and overcome the limitations of slow network access. However, as storage requirements tend to grow much faster than compute requirements, node scaling within HDFS tends to lead to wastage of compute resources.

An example of an issue caused by this would be that if HDFS were to have to store 10 petabytes of data due to its replication factor of 3, it would need to store 30 petabytes on the whole. At a max storage of 100TB per node, this would need 300 nodes, which would clearly overprovision compute facilities, thus causing other overheads. 

Minio overcomes these issues by the natural solution of separating storage an compute resources. Along with its cloud-native infrastructure, it is able to use orchestration frameworks like Kubernetes, the software stack.

\subsubsection{Performance}
To compare the performance of the two filesystems, an experiment is set up. First, the infrastructure is benchmarked to see the baseline limitations of the setup.
\begin{itemize}
    \item Hard Drive Performance
        \begin{itemize}
            \item Write: 137 MB/s
            \item Read: 205 MB/s
            \item Multi-drive performance with 32 threads + 32kb blocks:
                \begin{itemize}
                    \item Write: 655 MB/s
                    \item Read: 1.26 GB/s
                \end{itemize}
        \end{itemize}
    \item Network Performance
        \begin{itemize}
            \item Ethernet cables support 3.125 GB/s, but with multiple connections, the sustained throughput was at around 1.26 GB/s
        \end{itemize}
\end{itemize}

After the benchmarking of the basic infrastructure, some degree of tuning is done for each of the filesystems. This, on the whole, is to ensure that each of the filesystems is using all the resources allocated to them to the maximum extent possible, as would be done in a normal stress test.

\begin{itemize}
    \item Minio
        \begin{itemize}
            \item 1.2 TB aggregate memory across 12 nodes. Tuning was done such that MapReduce jobs could use the whole allocated CPU and Memory provided by compute nodes.
            \item The entire 144GB of RAM of each node was used.
            \item S3A connector (API) is tuned
        \end{itemize}
    \item HDFS
        \begin{itemize}
            \item Tuned to 2.4 TB aggregate memory across 12 nodes. 
            \item Tuned til the entirety of the 256 GB Ram on all compute nodes was being used. (Higher RAM due to shared compute and storage nodes).
            \item Ensured that computations do not go into swap space as well (as that causes lower performance)
            \item Configured to replicate data with a factor of 3
        \end{itemize}
\end{itemize}

The following tasks, which are considered to be Hadoop's most proven benchmarks, are evaluated:
\begin{itemize}
    \item Terasort
    \item Sort
    \item Wordcount
\end{itemize}

\subsection{\textbf{HDFS vs GFS}}
Sources \cite{lecturehdfsvsgfs} and data published in the papers of HDFS
and GFS are used to draw comparisons.

GFS was built for the unique of Google as a company: The use of off-the-shelf
hardware to run production servers, their scale of batch data processing, the
generally append only the nature of their high-throughput, latency-insensitive
workload. 

On the other hand, HDFS was implemented for the purpose of
running Hadoop’s MapReduce applications. It was created as an open-source
framework for the usage of different clients with different needs. This makes
Hadoop is far less opinionated; Therefore, it is also far less optimized at handling
workloads that GFS is tuned for.

In terms of data storage, recall that GFS chunks data into 64 MB chunks that
are uniquely identified. These chunks are replicated into 64 KB blocks
which are checksummed. This permits fast chunked reads while allowing
for error detection using the per-block checksum. On the other hand, HDFS 
divides data into  128MB blocks. The HDFS NameNode holds block replica
as two files: one with the data, the other with the checksum and generation stamp.

In terms of architecture, GFS is more involved due to its focus on
concurrent, atomic appends and snapshotting support. In particular, GFS
requires leases. The client is told where to write by the master. In HDFS,
the client decides where to write.

Data about reads and writes are collated from the data published by Google
on GFS performance as it runs on real-world clusters (Cluster B in ) \cite{ghemawat2003google}).

\begin{table}[H]
\centering
\label{tab:hdfsvsgfs}
 \caption{performance comparison: HDFS v/s GFS on production workloads. HDFS data gathered using 3000 nodes on DFSIO
 test suite. GFS numbers from Google's reported performance in production}
\begin{tabular}{c c c}
\hline
 \bf Measurement &  \bf GFS &  \bf HDFS \\
 \hline
 \it Read (Busy)  & 380 MB/s &  1.02 MB/s/node \\
 \it Writes (Busy) & 117 MB/s  & 1.09 MB/s per node \\
 \end{tabular}
\end{table}

\section{Conclusion}
In this paper, a detailed evaluation of three prominent distributed file systems (DFSs)—Google File System (GFS), Hadoop Distributed File System (HDFS), and MinIO—focusing is conducted on their scalability, fault tolerance, and overall performance in large-scale, dynamic environments. Each of these systems brings unique advantages and challenges to managing vast datasets across distributed networks, with varying approaches to data redundancy, server failures, and client access protocols.

The analysis has demonstrated that fault tolerance is a critical factor for ensuring consistent data availability and integrity in the face of network or server failures. While GFS and HDFS employ replication and redundancy techniques to maintain data availability, MinIO offers a more lightweight and flexible approach, catering to cloud-native environments. Scalability, as explored, is another defining feature of these systems, with all three DFSs leveraging mechanisms such as sharding and multi-node architectures to scale efficiently in response to increasing data loads and client demands.

Additionally, the impact of system design on performance was assessed, revealing that the choice of DFS is highly dependent on the specific enterprise requirements—whether prioritizing high availability, fault tolerance, or ease of integration with cloud computing infrastructures. For organizations involved in big data analytics, both HDFS and GFS offer robust, well-established solutions, while MinIO provides an appealing alternative for cloud-centric, object-based storage.

In conclusion, selecting the most appropriate DFS depends on the scale of operations, fault tolerance requirements, and integration needs with other cloud-based or distributed computing services. This paper serves as a comprehensive guide for understanding the strengths, limitations, and suitability of these DFSs, providing valuable insights for enterprises seeking to optimize their data management strategies in large-scale distributed environments.

\nocite{*} 
\bibliographystyle{IEEEtran}
\bibliography{references}
\end{document}